\documentstyle[12pt,aaspp4,flushrt,psfig]{article}
\def\etal{{\it et al.\ }}
\def\eg{{\it e.g.}}

\def\mnras{{\it Mon. Not. R. Astron. Soc.\ }}

\def\grays{$\gamma$-rays\ }
\def\apj{{\it Astrophys. J.\ }}
\def\prl{{\it Phys. Rev. Letters\ }}
\def\nat{{\it Nature\ }}

\begin{document}

\title{IMPLICATIONS OF ULTRAHIGH ENERGY AIR SHOWERS FOR PHYSICS AND 
ASTROPHYSICS}

\author{F.W. STECKER}

\affil{Laboratory for High Energy Astrophysics\\
NASA Goddard Space Flight Center, Greenbelt, MD, USA}

\begin{abstract}
    The primary ultrahigh energy particles which produce giant extensive air 
showers in the Earth's atmosphere present an intriguing mystery from two 
points of view: (1) How are these particles produced with such astounding 
energies, eight orders of magnitude higher than those produced by the best
man-made terrestrial accelerators? (2) Since they are most likely 
extragalactic in origin, how do they reach us from extragalactic distances
without suffering the severe losses expected from interactions with the 2.7 K
thermal cosmic background photons -- the so-called GZK effect?

    The answers to these questions may involve new physics:  violations of 
special relativity, grand unification theories, and quantum gravity theories 
involving large extra dimensions. They may involve new astrophysical sources,
"zevatrons". Or some heretofore totally unknown physics or astrophysics may
hold the answer. I will discuss here the mysteries involving the 
production and extragalactic propagation of ultrahigh energy cosmic rays 
and some suggested possible solutions.

\end{abstract}

\keywords{ultrahigh energy cosmic rays, active galactic nuclei, gamma-ray 
bursts, topological defects, grand unification}

\section{Introduction}

About once per century per km$^2$ of the Earth's surface, a giant shower of 
charged particles produced by a primary particle with an energy greater than 
or equal to 16 joules (100 EeV = $10^{20}$ eV) plows through the Earth's 
atmosphere. The showers which they produce can be detected by arrays of 
scintillators on the ground; they also announce their presence by producing a 
trail of ultraviolet flourescent light, exciting the nitrogen atoms in the 
atmosphere. The existence of such showers has been known for almost four
decades \cite{li63} (Linsley 1963). The number of giant air showers detected 
from primaries of energy greater than 100 EeV has grown into the double digits and may grow into the hundreds as new detectors such as the 
``Auger'' array and the ``EUSO'' (Extreme Universe Space Observatory) 
and ``OWL'' (Orbiting Wide-Angle Light Collectors) satellite detectors 
come on line. These phenomena present an intriguing 
mystery from two points of view: (1) How are particles produced with such 
astounding energies, eight orders of magnitude higher than are produced by 
the best man-made terrestrial accelerators? (2) Since they are most likely 
extragalactic in origin, how do they reach us from extragalactic distances 
without exhibiting the predicted cutoff from interactions with the 2.7K cosmic 
background radiation? In these lectures, I will consider possible solutions 
to this double mystery.

\section{The Data}

Figure 1 shows the published data (as of this writing) on the ultrahigh 
energy cosmic ray spectrum from the Fly's Eye and AGASA 
detectors.\footnote{The AGASA data have been reanalysed and the number of 
events determined to be above 100 EeV has been lowered to eight. (Teshima,
private communication.)} Other data from Havera Park and
Yakutsk may be found in the review by \cite{na00} Nagano and Watson (2000) 
are consistent with Figure 1. Additional data are now being obtained by the 
HiRes detector array and should be available in the near future (T. 
Abu-Zayyad, \etal , in preparation).

\begin{figure}
\centerline{\psfig{figure=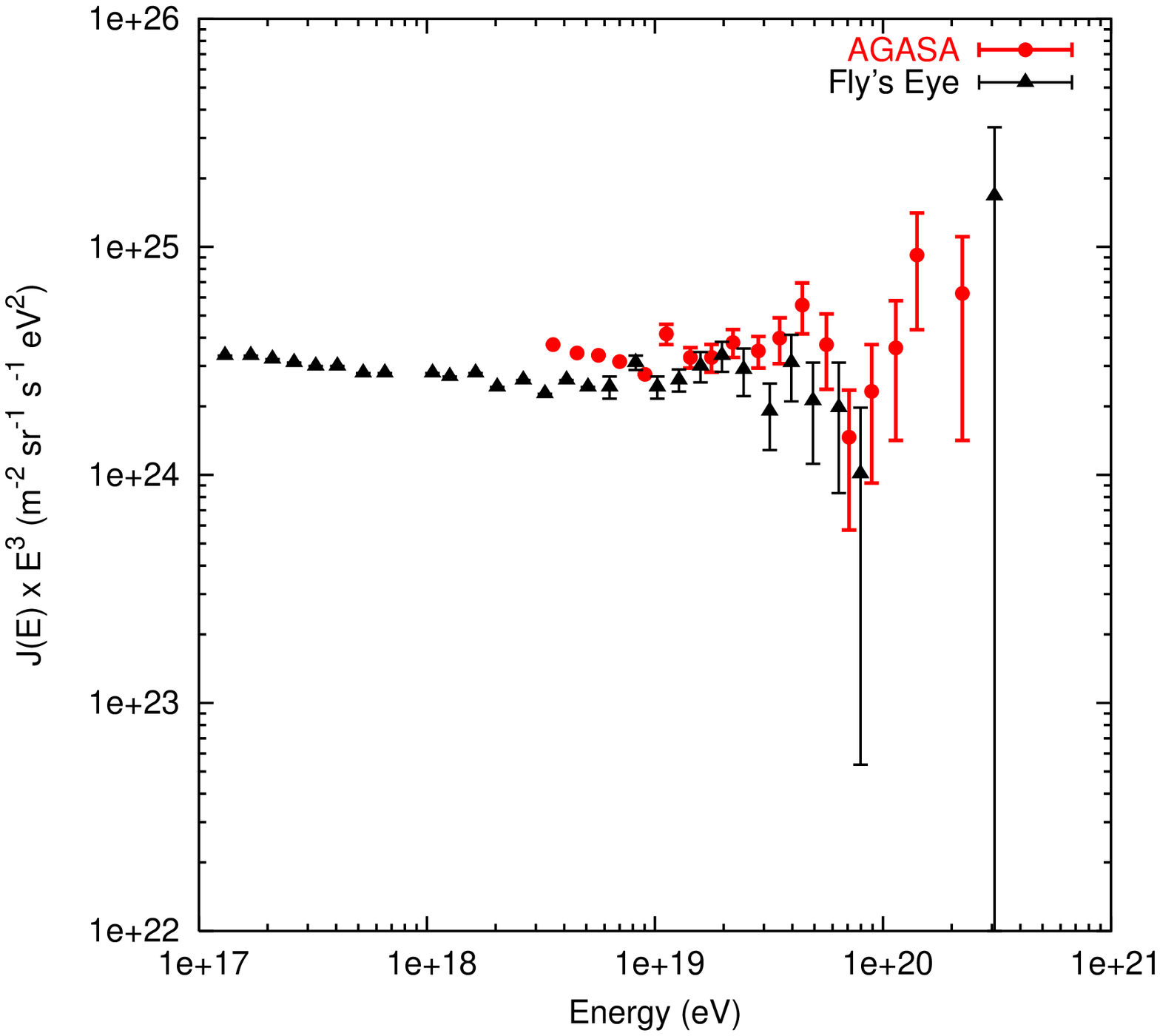,height=9cm}}

\caption{The ultrahigh energy cosmic ray spectrum data from Fly's Eye and
AGASA.}

\end{figure}

For air showers produced by primaries of energies in the 1 to 3 EeV range, 
\cite{ha99} Hayashida, {\it et al.} (1999) have found a marked directional 
anisotropy 
with a 4.5$\sigma$ excess from the galactic center region, a 3.9$\sigma$ 
excess from the Cygnus region of the galaxy, and a 4.0$\sigma$
deficit from the galactic anticenter region. This is strong evidence that
EeV cosmic rays are of galactic origin. A galactic plane enhancement in
EeV events was also reported by the Fly's Eye group (Dai, \etal 1999).
\cite{da99} 

As shown in Figure 2, at EeV energies, the primary particles
appear to have a mixed or heavy composition, trending toward a light 
composition in the higher energy range around 30 EeV \cite{bi93} 
(Bird, {\it et al.} 1993; \cite{ab00}
\cite{ab00} Abu-Zayyad, {\it et al.} 2000). This trend, together with 
evidence of a flattening in the cosmic ray spectrum on the 3 to 10 EeV energy 
range \cite{bi94}(Bird, {\it et al.} 1994; \cite{ta98} 
Takeda {\it et al.} 1998) is evidence for
a new component of cosmic rays dominating above 10 EeV energy. 

The apparent isotropy (no galactic-plane enhancement) of cosmic rays above
10 EeV \cite{ta99} ({\it e.g.}, Takeda, {\it et al.} 1999), together with
the difficulty of confining protons in the galaxy at 10 to 30 EeV energies,
provide significant reasons to believe that the cosmic-ray component above 
10 EeV is extragalactic in origin. As can be seen from Figure 1, this
extragalactic component appears to extend to an energy of 300 EeV.
Extention of this spectrum to higher energies is conceivable because such 
cosmic rays, if they exist, would be too rare to have been seen with
present detectors. We will see in the next section that the existence of 300 
EeV cosmic rays gives us a new mystery to solve.

\begin{figure}
\centerline{\psfig{figure=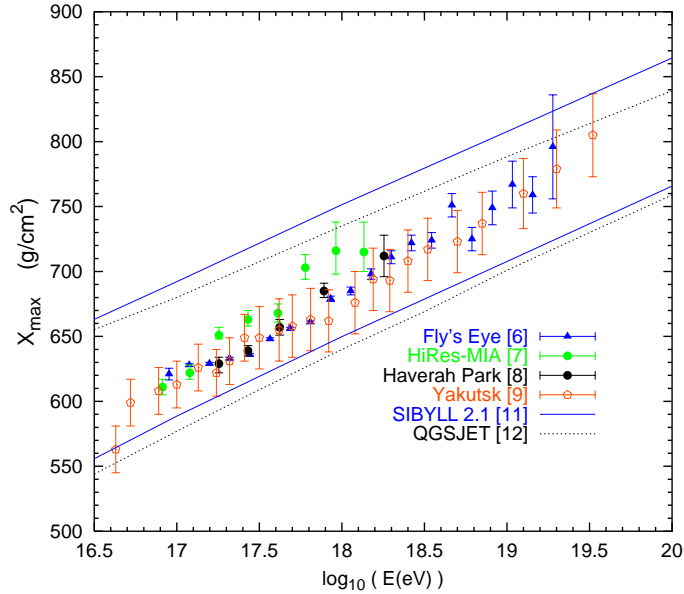,height=8cm}}

\caption{Average depth of shower maximum ($X_{max}$) {\it vs.} energy
compared to the calculated values for protons (upper curves) and Fe
primaries (lower curves) (from \protect\cite{ga00} Gaisser 2000; 
see references therein).}

\end{figure}

\section{The GZK Effect}

Thirty seven years ago, Penzias and Wilson (1965) \cite{pe65} reported the 
discovery of
the cosmic 3K thermal blackbody radiation which was produced very
early on in the history of the universe and which led to the undisputed
acceptance of the ``big bang'' theory of the origin of the universe. Much
more recently, the Cosmic Background Explorer (COBE) satellite confirmed this
discovery, showing that the cosmic background radiation (CBR) has the 
spectrum of the most perfect thermal blackbody known to man. COBE data also 
showed that this radiation (on angular scales $ > 7^\circ$) was isotropic to a 
part in $10^5$ \cite{mat94}(Mather {\it et al.} 1994). The perfect 
thermal character and smoothness of the CBR proved
conclusively that this radiation is indeed cosmological and that, at the 
present time, it fills the entire universe with a 2.725 K thermal
spectrum of radio to 
far-infrared photons with a density of $\sim 400$ cm$^{-3}$.

\begin{figure}
\centerline{\psfig{figure=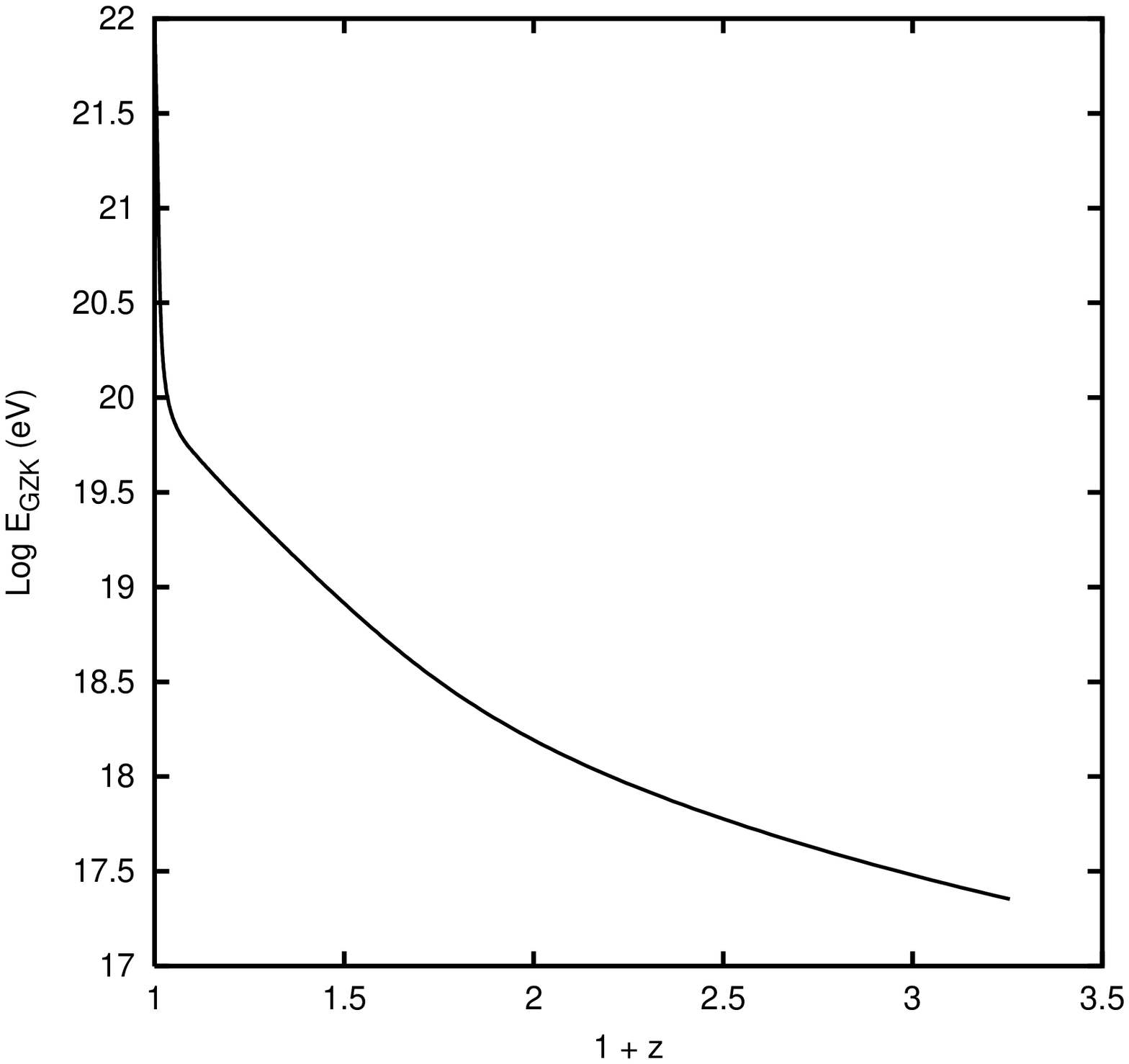,height=9cm}}

\caption{The GZK cutoff energy versus redshift \protect\cite{sc02} (Scully 
and Stecker 2002).}

\end{figure}

Shortly after the discovery of the CBR, \cite{gr66} Greisen (1966) and 
Zatsepin and Kuz'min (1966)\cite{za66} 
predicted that pion-producing interactions of ultrahigh energy cosmic 
ray protons with CBR photons of target density $\sim$ 400 cm$^{-3}$ should 
produce a cutoff in their spectrum at energies greater than $\sim$ 50 EeV. 
This predicted effect has since become known as the GZK 
(Greisen-Zatsepin-Kuz'min) effect. Following the GZK papers, Stecker (1968)
\cite{st68} utilized data on the energy dependence of the 
photomeson production cross 
sections and inelasticities to calculate the mean energy loss time for protons
propagating through the CBR in intergalactic space as a function of energy.
Based on his results, Stecker (1968) then suggested that the particles of 
energy above the GZK cutoff energy (hereafter referred to as trans-GZK 
particles) must be coming from within the ``Local Supercluster'' of which we 
are a part and which is centered on the Virgo Cluster of galaxies. Thus, the 
``GZK cutoff'' is not a true cutoff, but a supression of the ultrahigh energy
cosmic ray flux owing to a limitation of the propagation distance to a few
tens of Mpc.

The actual position of the GZK cutoff can differ from the 50 EeV predicted
by Greisen. In fact, there could actually be an {\it enhancement} at or near 
this energy owing to a ``pileup'' of cosmic rays starting out at higher 
energies and crowding up in energy space at or below the predicted cutoff 
energy \cite{pu76}(Puget, Stecker and Bredkamp 1976; \cite{hi85} Hill and 
Schramm 1985; \cite{be88} Berezinsky and Grigor'eva 1988; \cite{st89} 
Stecker 1989; \cite{st99} Stecker and Salamon 1999).
The existence and intensity of this predicted pileup depends critially on the
flatness and extent of the source spectrum, ({\it i.e.}, the number of cosmic 
rays starting out at higher energies), but if its existence is confirmed in the
future by more sensitive detectors, it would be evidence for the GZK effect.

Scully and Stecker (2002) have determined the GZK energy, defined as the
energy for a flux decrease of $1/e$, as a function of redshift. At high
redshifts, the target photon density increases by $(1+z)^3$ and both the 
photon and initial cosmic ray energies increase by $(1+z)$. The results
obtained by Scully and Stecker are shown in Figure 3.

\section{Acceleration and Zevatrons: The ``Bottom Up'' Scenario}

\begin{figure}
\centerline{\psfig{figure=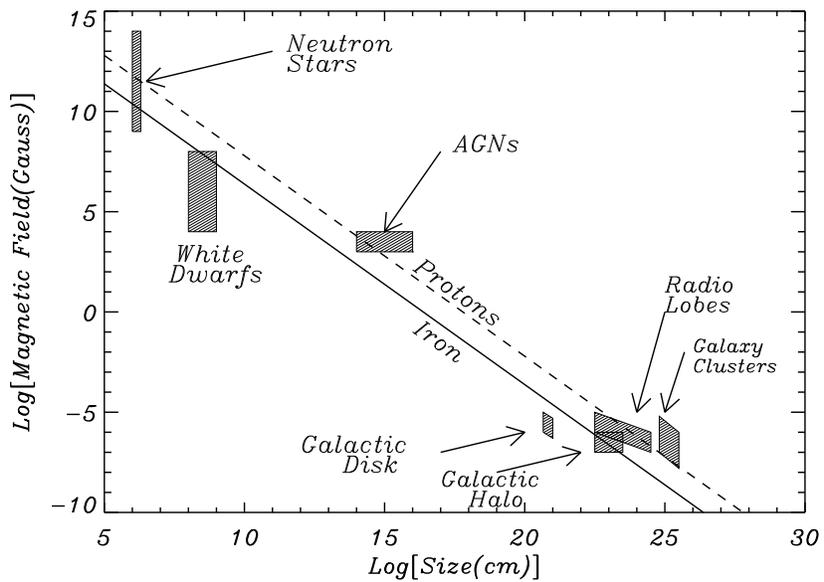,height=8cm}}

\caption{A ``Hillas Plot'' showing potential astrophysical zevatrons 
(from Olinto 2000).\protect\cite{ol00} The lines are for $B$ {\it vs.} $L$ for
$E_{max}$ = 0.1 ZeV for protons and iron nuclei as indicated.}

\end{figure}

The apparent lack of a GZK cutoff has led theorists to go on a hunt for nearby
``zevatrons'', {\it i.e.}, astrophysical sources which can accelerate 
particles to energies $\cal{O}$(1 ZeV = 10$^{21}$eV).

In most theoretical work in cosmic ray astrophysics, it is generally assumed 
that the diffusive shock acceleration process is the most likely mechanism for
accelerating particles to high energy. (See, {\it e.g.}, \cite{jo00a} Jones 
(2000) and references therein.) In this case, the maximum obtainable
energy is given by $E_{max}=keZ(u/c)BL$, where $u \le c$ is the shock speed, 
$eZ$ is the charge of the particle being accelerated, $B$ is the magnetic field
strength, $L$ is the size of the accelerating region and the numerical 
parameter $k = \cal{O}$$(1)$ (\cite{dr94} Drury 1994). 
Taking $k = 1$ and $u = c$, one finds

$$ E_{max} = 0.9Z(BL) $$

\noindent with $E$ in EeV, $B$ in $\mu$G and $R$ in kpc. This assumes that 
particles can be accelerated efficiently up until the moment when they can 
no longer be contained by the source, {\it i.e.} until their gyroradius 
becomes larger
than the size of the source. Hillas (1984) \cite{hi84} used this relation to 
construct
a plot of $B$ {\it vs.} $L$ for various candidate astrophysical objects. A
``Hillas plot'' of this kind, recently constructed by Olinto (2000), is shown 
in Figure 4.
 
Given the relationship between $E_{max}$ and $BL$ as shown in Figure 4, there 
are not too many astrophysical candidates for zevatrons. Of these, galactic
sources such as white dwarfs, neutron stars, pulsars, and magnetars can be 
ruled out because their galactic distribution would lead to anisotropies above
10 EeV which would be similar to those observed at lower energies by
\cite{ha99} Hayashida \etal (1999), and this is not the case. 
Perhaps the most promising potential zevatrons are radio lobes of strong radio 
galaxies (Biermann and Strittmatter 1987) \cite{bi87}. The trick is that such 
sources need to be 
found close enough to avoid the GZK cutoff ({\it e.g.}, Elbert and Sommers 
1995)\cite{el95}. Biermann has further suggested
that the nearby radio galaxy M87 may be the source of the observed trans-GZK 
cosmic rays (see also Stecker 1968; \cite{fa00} Farrar and Piran 2000). Such 
an explanation would require one to invoke magnetic field
configurations capable of producing a quasi-isotropic distribution of 
$> 10^{20}$ eV protons, making this hypothesis questionable. However, if the 
primary particles are nuclei, it is easier to explain a radio galaxy
origin for the two highest energy events \cite{st99}(Stecker and Salamon 
1999; see section 4.3). 

\subsection{The Dead Quasar Origin Hypothesis}

It has been suggested that since all large galaxies are suspected to
harbor supermassive black holes in their centers which may have once been
quasars, fed by accretion disks which are now used up, that nearby quasar
remnants may be the searched-for zevatrons (Boldt and Ghosh 1999; Boldt
and Lowenstein 2000)\cite{bo99} \cite{bo00}. This scenario also has potential 
theoretical problems and needs to be explored further. In particular, it
has been shown that black holes which are not accreting plasma cannot
possess a large scale magnetic field with which to accelerate particles
to relativistic energies (Ginzburg and Ozernoi 1964; Krolik 1999;
Jones 2000). \cite{gi64} \cite{kr99} \cite{jo00}
Observational evidence also indicates that the cores of weakly active
galaxies have low magnetic fields (Falcke 2001 and references therein.) 
\cite{fa01}

\subsection{Gamma-Ray Burst Zevatrons} 

In 1995, it was hypothesized that cosmological $\gamma$-ray bursts (GRBs)
could be  the source of the highest energy cosmic rays (Waxman 1995; Vietri 
1995).\cite{wa95} \cite{vi95} It was suggested that if 
these objects emitted the same amount of energy in
ultrahigh energy ($\sim 10^{14}$ MeV) cosmic rays as in $\sim$ MeV photons, 
there would be enough energy input of these particles into intergalactic 
space to account for the observed flux. At that time, it was assumed that the 
GRBs were distributed uniformly, independent of redshift. 

\subsubsection{Cosmological GRBs and the GZK Problem}

In recent years, X-ray, optical, and radio afterglows of about a dozen
GRBs have been detected leading to the subsequent identification of the host
galaxies of these objects and consequently, their redshifts. 
To date, some 14 GRBs afterglows have been detected with a subsequent
identification of their host galaxies.  As of this writing, 13 of the 14 are 
at moderate to high redshifts with the highest one (GRB000131) lying at a 
redshift of 4.50.

A good argument in favor of strong redshift evolution for the frequency 
of occurrence of the higher luminosity GRBs has been made by Mao and Mo 
(1998)\cite{ma98}, based on the star-forming nature of the host galaxies. 
The host galaxies of GRBs appear to be sites of active star
formation. The colors and morphological types of the host
galaxies are indicative of ongoing star formation, as
is the detection of Ly$\alpha$ and [OII] in several of these galaxies. 
Further evidence suggests that bursts themselves are directly associated with 
star forming regions within their host galaxies; their positions
correspond to regions having significant hydrogen column densities
with evidence of dust extinction. 
Results of the analysis of Schmidt (1999).\cite{sc99} also favors a GRB 
redshift distribution which follows the strong redshift evolution of the 
star formation rate. Thus, it now seems reasonable to assume
that a more appropriate redshift
distribution to take for GRBs is that of the average star
formation rate, rather than a uniform distribution.

If we thus assume a redshift distribution for the GRBs which follows the star
formation rate, being significantly higher at higher redshifts, 
GRBs fail by at least an order of magnitude to account
for the observed cosmic rays above 100 EeV (Stecker 2000)\cite{st00}. 
If one wishes to account for the GRBs above 10 EeV, this hypothesis fails by 
two to three orders of magnitude (Scully and Stecker 2002)\cite{sc02}. Even 
these numbers are most likely too optimistic, since they are based on the 
questionable assumption of the same amount of GRB
energy being put into ultrahigh energy cosmic rays as in $\sim$ MeV photons.

Figure 5, from Scully and Stecker, (2002)\cite {sc02} shows the {\it form} of 
the cosmic 
ray spectrum to be expected from sources with a uniform redshift distribution 
and sources which follow the star formation rate. The required normalization 
and spectral index determine the energy requirements of any cosmological 
sources which are invoked to explain the observations. Pileup effects and GZK 
cutoffs are evident in the theoretical curves in this figure. As can be seen 
in Figure 5, the present data appear to be statistically consistent with either
the presence or absence of a pileup effect. Future data with much better
statistics are required to determine such a spectral structure.

\begin{figure}
\centerline{\psfig{figure=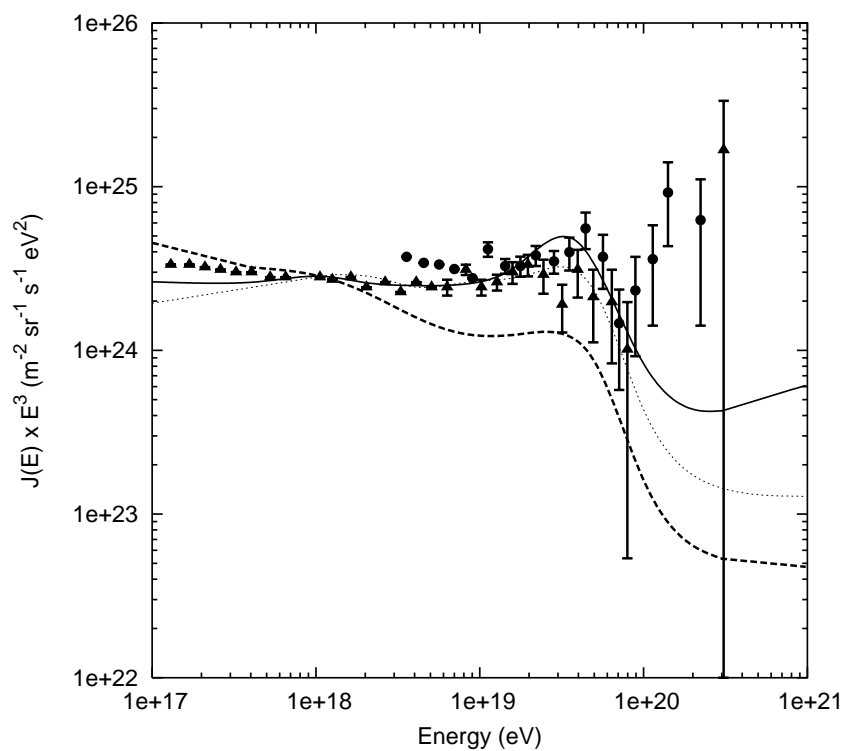,height=16cm}}

\caption{Predicted spectra for cosmic ray protons as compared with the data.
The middle curve and lowest curve assume an $E^{-2.75}$ source spectrum with
a uniform source distribution and one that follows the $z$ distribution of the
star formation rate respectively. The upper curve is for an $E^{-2.35}$ source
spectrum which requires an order of magnitude more energy input and exhibits
a ``pileup effect'' (Scully and Stecker 2002).\protect\cite{sc02}}

\end{figure}

\subsubsection{Low Luminosity Gamma Ray Bursts}

An unusual nearby Type Ic supernova, SN 1998bw, has been identified as the 
source of a low luminosity burst, GRB980425, with an energy release 
which is orders of magnitude smaller than that for a typical cosmological GRB.
Norris (2002) \cite{no02} has given an analysis of the luminosities and
space densities of such nearby low luminosity long-lag GRB sources which
are identified with Type I supernovae. For these sources, he finds a rate
per unit volume of $7.8 \times 10^{-7}$ Mpc$^{-3}$yr$^{-1}$ and an average
(isotropic) energy release per burst of 1.3 $\times 10^{49}$ erg over the
energy range from 10 to 1000 keV. The energy release per unit volume is then
$\sim 10^{43}$ erg Mpc$^{-3}$yr$^{-1}$. This rate is more than an order of
magnitude below the rate needed to account for the cosmic rays with energies
above 10 EeV.

\subsection{The Heavy Nuclei Origin Scenario}

\begin{figure}
\centerline{\psfig{figure=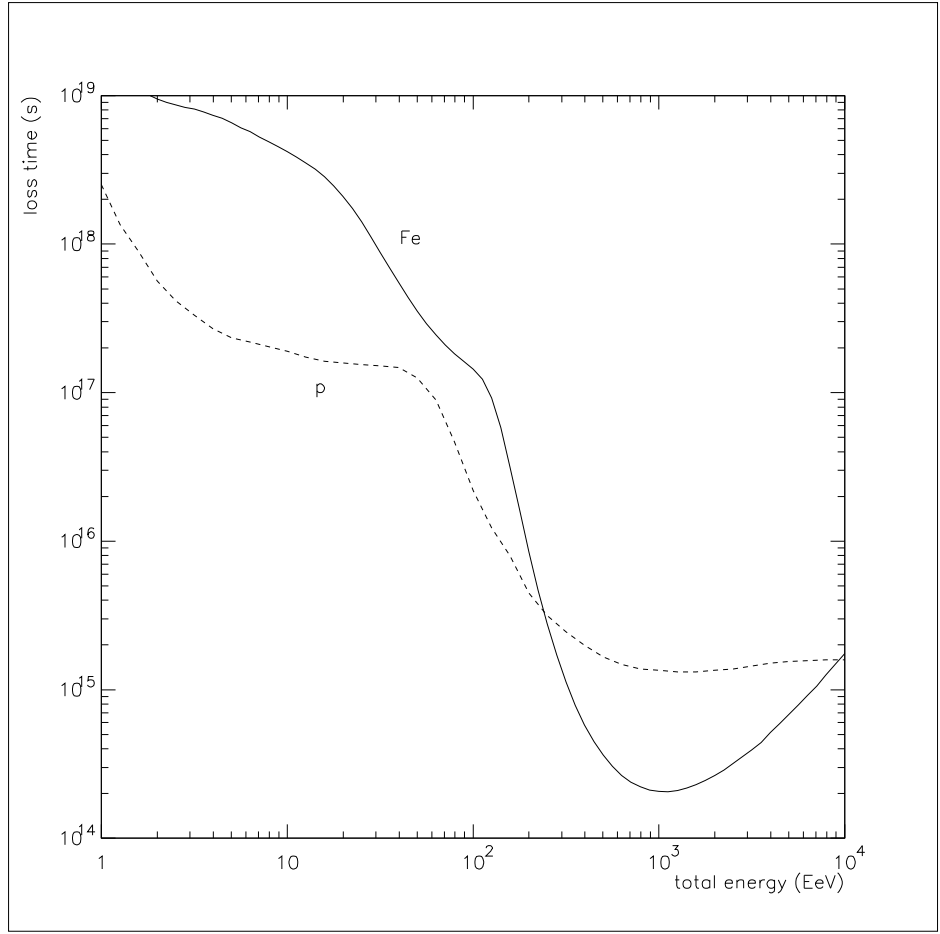,height=16cm}}

\caption{Mean energy loss times for protons (Stecker 1968; Puget, Stecker and
Bredekamp 1976) and nuclei originating as Fe (Stecker and Salamon 1999).}

\end{figure}

A more conservative hypothesis for explaining the trans-GZK events is that they
were produced by heavy nuclei. Stecker and Salamon (1999) have shown that the
energy loss time for nuclei starting out as Fe is longer than that for protons
for energies up to a total energy of $\sim$300 EeV (see Figure 6). 

Stanev \etal (1995) and Biermann (1998) have examined the arrival directions 
of the highest energy events.\cite{st95} \cite{Bi98} They point out 
that the $\sim 200$ EeV event is within 10$^\circ$ of the direction of the 
strong radio galaxy NGC 315. This galaxy lies at a distance of only 
$\sim$ 60 Mpc from us. For that distance, the results of Stecker and Salamon 
(1999) indicate that heavy nuclei would
have a cutoff energy of $\sim$ 130 EeV, which may be within the uncertainty in
the energy determination for this event. The $\sim$300 EeV event is within
12$^\circ$ of the direction of the 
strong radio galaxy 3C134. The distance to 3C134 
is unfortunately unknown because its location behind a dense molecular cloud 
in our own galaxy obscures the spectral lines required for a measurement of 
its redshift. It may be possible that {\it either} 
cosmic ray protons {\it or} heavy nuclei originated in 
these sources and produced the highest energy air shower events.

An interesting new clue that we may indeed be seeing heavier nuclei above the
proton-GZK cutoff comes from a recent analysis of inclined air showers
above 10 EeV energy (Ave, {\it et al.} 2000). These new results favor
proton primaries below the p-GZK cutoff energy but they {\it appear to favor a 
heavier composition above the p-GZK cutoff energy}. It will be interesting to
see what future data from much more sensitive detectors will tell us.

\section{Top-Down Scenarios: ``Fraggers''}

A way to avoid the problems with finding plausible astrophysical zevatrons is
to start at the top, {\it i.e.}, the energy scale associated with grand
unification, supersymmetric grand unification or its string theory equivalent.

  The modern scenario for the early history of the big bang takes account of 
the work of particle theorists to unify the forces of nature in the framework 
of Grand Unified Theories (GUTs) (\eg , Georgi and Glashow 1974). \cite{ge74} 
This concept extends the very successful work of Nobel Laureates Glashow, 
Weinberg, and Salam in unifying the electromagnetic and weak nuclear forces 
of nature (Glashow 1960; Weinberg 1967; Salam 1968). \cite{gl60} \cite{we67}
\cite{sa68}  
As a consequence of this theory, the electromagnetic and weak forces would
have been unified at a higher temperature phase in the early history 
of the universe and then would have been broken into separate forces through
the mechanism of spontaneous symmetry breaking caused by vacuum fields which 
are known as Higgs fields.

In GUTs, this same paradigm is used to infer that the electroweak force
becomes unified with the strong nuclear force at very high 
energies of $\sim 10^{24}$ eV 
which occurred only $\sim 10^{-35}$ seconds 
after the big bang. The forces then became separated owing to interactions
with the much heavier mass scale Higgs fields whose symmetry was broken
spontaneously. The supersymmetric GUTs (or SUSY GUTs) provide an explanation
for the vast difference between the two unification scales (known as the
``Hierarchy Problem'') and predict that the running coupling constants 
which describe the strength of the various forces become equal at the SUSY GUT
scale of $\sim 10^{24}$ eV (Dimopoulos, Raby and Wilczek 1982). \cite{di82}

\subsection{Topological Defects: Fossils of the Grand Unification Era}

The fossil remnants of this unification are predicted to be very heavy 
topological defects in the vacuum of
space caused by misalignments of the heavy Higgs fields in regions which
were causally disconnected in the early history of the universe. These are 
localized regions where extremely high densities of mass-energy are trapped. 
Such defects go by designations such as cosmic strings, monopoles, walls,
necklaces (strings bounded by monopoles), and textures, depending on their 
geometrical and topological properties.  Inside a topological defect 
vestiges of the early universe may be preserved to the present day.  
The general scenario for creating topological defects in the early universe 
was suggested by Kibble (1976)\cite{ki76}.

Superheavy particles or topological structures arising at the GUT energy scale
$M \ge 10^{23}$ eV can decay or annihilate to produce ``X-particles'' (GUT 
scale Higgs particles, superheavy fermions, or leptoquark bosons of mass M.) 
In the case of strings
this could involve mechanisms such as intersecting and intercommuting
string segments and cusp evaporation. These X-particles will 
decay to produce QCD fragmentation jets at ultrahigh energies, so I will 
refer to them as ``fraggers''. QCD fraggers produce mainly pions, with a 3 to 
10 per cent admixture of
baryons, so that generally one can expect them to produce at least an order of 
magnitude more high energy $\gamma$-rays and neutrinos than protons. 
The same general scenario would hold for the decay of long-lived superheavy 
dark matter particles (see section 5.3), which would also be fraggers. It has 
also been suggested that the decay of ultraheavy particles from topological 
defects produced in SUSY-GUT models which can
have an additional soft symmetry breaking scale at TeV energies (``flat SUSY
theories'') may help explain the observed $\gamma$-ray background flux at
energies $\sim$ 0.1 TeV (Bhattacharjee, Shafi and Stecker 1998)\cite{bh98}. 

The number of variations and models for explaining the ultrahigh energy
cosmic rays based on the GUT or SUSY GUT scheme (which have come to be
called ``top-down'' models) 
has grown to be enormous and I will not attempt to list all of the
numerous citations involved. Fortunately, Bhattacharjee and Sigl (2000) 
\cite{bh00} have
recently published an extensive review with over 500 citations and I refer the
reader to this review for further details of ``top-down'' models and 
references. The important thing to note here
is that, if the implications of such models are borne out by future cosmic
ray data, they may provide our first real evidence for GUTs.

\subsection{``Z-bursts''}

It has been suggested that ultra-ultrahigh energy $\cal{O}$(10 ZeV) neutrinos
can produce ultrahigh energy $Z^0$ fraggers by interactions with  1.9K thermal
CBR neutrinos (Weiler 1982; Fargion, \etal 1999; Weiler 1999), \cite{fa99}
\cite{we99} resulting in ``Z-burst'' fragmentation jets, again producing 
mostly pions. This will occur at the resonance energy $E_{res} =
4[m_{\nu}({\rm eV})]^{-1}$ ZeV. A typical $Z$ boson will decay to produce
$\sim$2 nucleons, $\sim$20 $\gamma$-rays and $\sim$ 50 neutrinos, 2/3 of 
which are $\nu_{\mu}$'s. 

If the nucleons which are produced from Z-bursts originate within a few tens of
Mpc of us they can reach us, even though the original $\sim$ 10 ZeV 
neutrinos could have come from a much further distance. 
It has been suggested that this effect can be amplified if our galaxy has
a halo of neutrinos with a mass of tens of eV (Fargion, Mele and Salis 1999;
Weiler 1999). \cite{fa99} \cite{we99} However, a neutrino mass large enough 
to be confined to a galaxy size neutrino halo (Tremaine and Gunn 1979) 
would imply a hot dark matter cosmology which is 
inconsistent with simulations of galaxy formation and clustering 
({\it e.g.}, Ma and Bertschinger 1994) \cite{ma94}
and with angular fluctuations in the CBR. (Another problem with halo
fraggers is discussed below in section 5.4) A mixed dark matter model with 
a lighter neutrino mass (Shafi and Stecker 1984) \cite{sh84} 
produces predicted CBR angular fluctuations (Schaefer, Shafi and Stecker 
1989) \cite{sc89} which are consistent with the {\it Cosmic Background 
Explorer} data (Wright 1992)\cite{wr92}. In such a model, neutrinos would 
have density fluctuations on the 
scale of superclusters, which would still allow for some amplification 
(Weiler 1999) \cite{we99a}. The tritium decay spectral endpoint
limits on the mass of the electron neutrino (Weinheimer, \etal 1999), 
\cite{we99} together with the very small neutrino flavor mass differences 
indicated by the atmospheric and solar neutrino oscillation results
(Ahmad, \etal 2002) \cite{ah02} constrains all neutrino flavors to have 
masses in the range $\cal{O}$(eV) or less. This is much too small a mass 
for neutrinos to to be confined to halos of individual galaxies.

The basic general problem with the Z-burst explanation
for the trans-GZK events is that one needs to produce 10 ZeV neutrinos. If 
these are secondaries from pion production, this implies that the primary
protons which produce them must have energies of hundreds of ZeV! Since we
know of no astrophysical source which would have the potential of accelerating
particles to energies even an order of magnitude lower (see section 4),
a much more likely scenario for producing 10 ZeV neutrinos would be by a 
top-down process. The production rate of neutrinos from such processes is
constrained by the fact that the related energy release into electromagnetic 
cascades which produce GeV range \grays is limited by the satellite 
observations (see the review by Bhattacharjee and Sigl 2000). \cite{bh00} 
This constraint, together with the low probability for 
Z-burst production, relegates the Z-burst phenomenon to a minor 
secondary role at best.

\subsection{Ultraheavy Dark Matter Particles: ``Wimpzillas''}

The idea has been suggested that the dark matter which makes up most of the
gravitating mass in the universe could consist of ultraheavy particles
produced by non-thermal processes in the early big-bang (Berezinsky \etal
1997; Kuz'min and Rubakov 1998; Blasi \etal 2002; Sarkar and 
Toldr\`{a} 2002; Barbot \etal 2002; see also the paper of Rocky Kolb in these 
proceedings.) \cite{be97} \cite{ku98} \cite{sa01} \cite{bl02} 
The annihilation or decay of such particles in a
dark matter halo of our galaxy would then produce ultrahigh energy nucleons
which would not be attenuated at trans-GZK energies owing to their proximity.

\subsection{Halo Fraggers and the Missing Photon Problem}

Halo fragger models such as Z-burst and ultraheavy halo dark matter 
(``wimpzilla'') decay or annihilation, as we have seen, will produce
more ultrahigh energy photons than protons. These ultrahigh energy photons 
can reach the Earth from anywhere in a dark matter galactic halo, because, as 
shown in Figure 7, there is a ``mini-window'' for the transmission of 
ultrahigh energy cosmic rays between $\sim 0.1$ and $\sim 10^{6}$ EeV.

Photon-induced giant air showers have an evolution profile which is 
significantly different from nucleon-induced showers because of the
Landau-Pomeranchuk-Migdal (LPM) effect \cite{la53} (Landau and Pomeranchuk
1953; Migdal 1956) \cite{mi56} 
and because of cascading in the Earth's magnetic field (Cillis, \etal 1999)
\cite{ci99} (see Figure 7). By taking this into account, Shinozaki, {\it
et al.} (2002) \cite{sh02} have used the AGASA data to place upper limits on 
the photon composition of their UHECR showers. They find a photon content 
upper limit of 28\% for events above 10 EeV and 67\% for events above 30 EeV 
at a 95\% confidence level with no indication of photonic showers above 100 
EeV. A recent reanalysis of the ultrahigh energy events observed at Haverah 
Park by Ave, {\it et al.} (2002) \cite{av02}
indicates that less than half of the events (at 95\% confidence level)
observed above 10 and 40 EeV are $\gamma$-ray initiated.
An analysis of the highest energy Fly's Eye event ($E = 300$ EeV) 
\cite{ha02} (Halzen and Hooper 2002) shows it not to be of photonic origin,
as indicated in Figure 8. In addition,
Shinozaki, {\it et al.} (2002) have found no indication of departures from
isotropy as would be expected from halo fragger photonic showers, this
admittedly with only 10 events in their sample.

\begin{figure}
\centerline{\psfig{figure=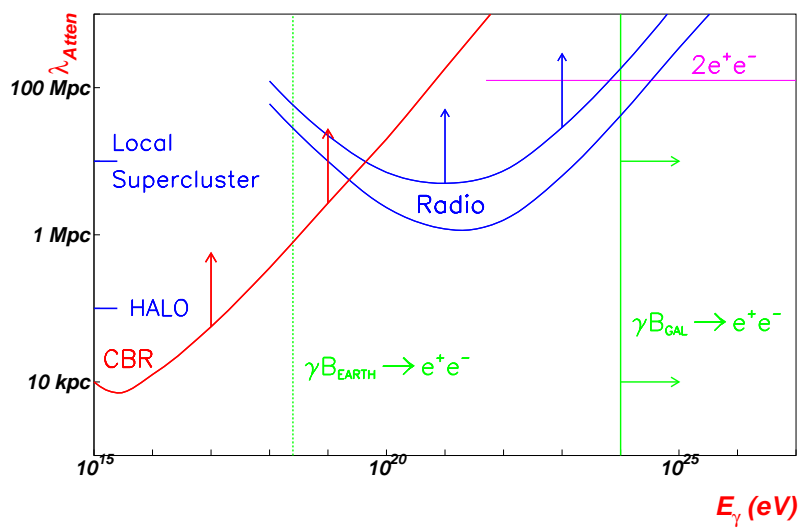,height=8cm}}

\caption{The mean free path for ultrahigh energy $\gamma$-ray attenuation
{\it vs.} energy. The curve for electron-positron pair production off the 
cosmic background radiation (CBR) is based on \protect\cite{go66} Gould and 
Schreder (1966). The two estimates for pair production off the extragalactic 
radio background are from \protect\cite{pr96} Protheroe and Biermann (1996).
The curve for double pair production is based on \protect\cite{br73} Brown, 
{\it et al.} (1973). The physics of pair production by single photons in 
magnetic fields is discussed by \protect\cite{er66} Erber (1966). This process
eliminates all photons above $\sim 10^{24}$ eV and produces a terrestrial
anisotropy in the distribution of photon arrival directions above 
$\sim 10^{19}$eV.}
\end{figure}

\begin{figure}
\centerline{\psfig{figure=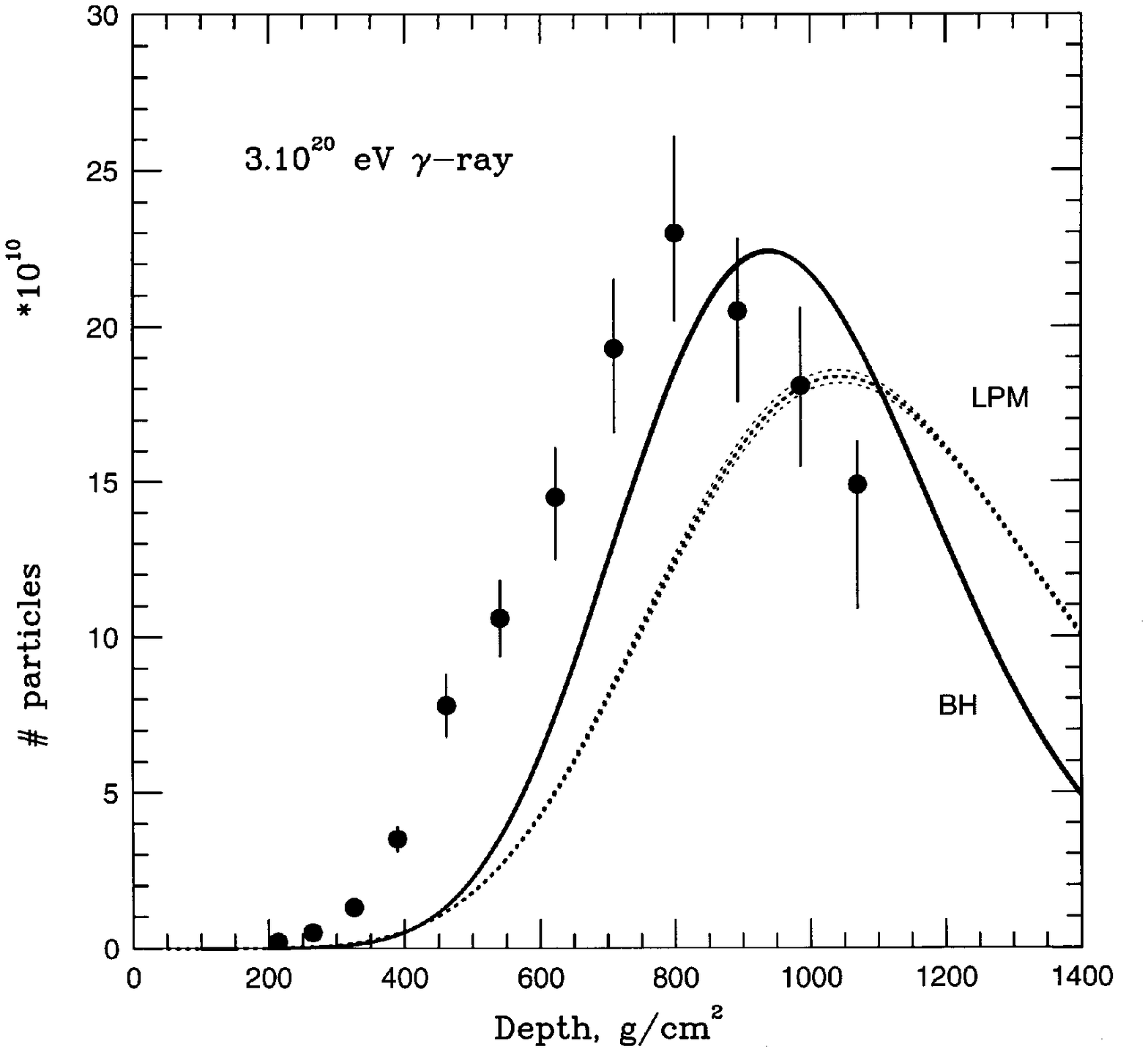,height=7cm}}

\caption{The composite atmospheric shower profile of a 300 EeV photon-induced
shower calculated with the Bethe-Heitler (solid) electromagnetic cross
section and with the LPM effect taken into account (dashed line, see text).
The measured Fly's Eye profile, which fits the profile of a nucleonic 
primary, is shown by the data points \protect\cite{ha02} (Halzen and Hooper 
2002).}
\end{figure}

\section{Other New Physics Possibilities}

The GZK cutoff problem has stimulated theorists to look for possible solutions
involving new physics. Some of these involve (A) a large increase in the
neutrino-nucleon cross section at ultrahigh energies, (B) new particles, 
and (C) a small violation of Lorentz Invariance (LI).

\subsection{Increasing the Neutrino-Nucleon Cross Section at Ultrahigh 
Energies}

Since neutrinos can travel through the universe without interacting with the
2.7K CBR, it has been suggested that if the neutrino-nucleon cross section 
were to increase to hadronic values at ultrahigh energies, they could produce 
the giant air showers and account for the observations of showers above the
proton-GZK cutoff. Several suggestions have been made for processes that can
enhance the neutrino-nucleon cross section
at ultrahigh energies. These suggestions include 
composite models of neutrinos
(Domokos and Nussinov 1987; Domokos and Kovesi-Domokos 1988)\cite{do87} \cite{do88}, scalar 
leptoquark  resonance channels (Robinett 1988) and the exchange of dual 
gluons (Bordes, {\it et al.} 
1998)\cite{bo98}. Burdman, Halzen and Ghandi (1998) \cite{bu98} have ruled out
a fairly general class of these types of models, including those listed above,
by pointing out that in order to increase the neutrino-nucleon cross section
to hadronic values
at $\sim 10^{20}$ eV without violating unitarity bounds, the relevant scale 
of compositeness or particle exchange would have to be of the order of a 
GeV, and that such a scale is ruled out by accelerator experiments.

More recently, the prospect of enhanced neutrino cross sections has been 
explored in the context of extra dimension models. Such models have been
suggested by theorists to unify the forces of physics since the days of
Kaluza (1921) \cite{ka21} and Klein (1926). \cite{kl26} In recent years,
they have been invoked by string theorists and by other theorists as a 
possible way for accounting for the extraordinary weakness of the gravitational
force, or, in other words, the extreme size of the Planck mass (Arkani-Hamed,
Dimopoulos and Dvali 1999; Randall and Sundrum 1999). \cite{ar99} \cite{ra99} 
These models allow the virtual exchange of gravitons propagating 
in the bulk ({\it i.e.} in the space of full extra dimensions) 
while restricting the propagation of other particles to the familiar four 
dimensional space-time manifold. It has been suggested that in such models,
$\sigma$($\nu$N) $\simeq [E_{\nu}/(10^{20}\rm eV)]$ mb (Nussinov and Schrock
1999; Jain, {\it et al.} 2000; see also Domokos and Kovesi-Domokos 1999). 
\cite{nu99} \cite{ja00} It should be noted that a cross section of
$\sim 100$ mb would be necessary to approach obtaining consistency with
the air shower profile data. Other scenarios involve the neutrino-initiated
atmopheric production of black holes (Anchordoqui, \etal 2002) \cite{an02}
and even higher dimensional extended objects, p-dimensional branes called 
``p-branes'' (Ahn, Cavalgia and Olinto 2002; Anchordoqui, Feng and Goldberg 
2002). \cite{ah02a} \cite{afg02} Such
interactions, in principle, can increase the neutrino total atmospheric 
interaction cross section by orders of magnitude above the standard model
value. However, as discussed by Anchordoqui, Feng and Goldberg (2002), 
\cite{afg02} sub-mm gravity experiments and astrophysical constraints rule
out total neutrino interaction cross sections as large as 100 mb as would be
needed to fit the trans-GZK energy air shower profile data. Nonetheless, extra
dimension models still may lead to significant increases in the neutrino
cross section, resulting in moderately penetrating air showers. Such 
neutrino-induced showers should also be present at somewhat lower energies and
provide an observational test for extra dimension TeV scale gravity models
(Anchordoqui, \etal 2001; Tyler, Olinto and Sigl 2001). \cite{an01} 
\cite{ty01} As of this writing, no such showers
have been observed, putting an indirect constraint on fragger scenarios with
TeV gravity models.

\subsection{New Particles}

The suggestion has also been made that new neutral particles containing
a light gluino could be producing the trans-GZK events (Farrar 1996;
Cheung, Farrar and Kolb 1998).\cite{fa96} \cite{ch98} While the invocation of such new particles
is an intriguing idea, it seems unlikely that such particles of a few
proton masses would be produced in copious enough quantities in astrophysical
objects without being detected in terrestrial accelerators. Also there
are now strong constraints on gluinos (Alavi-Harati, {\it et al.} 1999)\cite{al99}.
One should note that while it is true that the GZK threshold for such 
particles would be higher than that for protons, 
such is also the case for the more prosaic heavy nuclei
(see section 4.3). In addition, such neutral particles cannot be accelerated 
directly, but must be produced as secondary particles, making the energetics
reqirements more difficult.

\subsection{Breaking Lorentz Invariance}  

With the idea of spontaneous symmetry breaking in particle physics came the
suggestion that Lorentz invariance (LI) might be weakly broken at high energies
(Sato and Tati 1972). Although no real quantum theory of gravity exists, it 
was suggested that LI might be broken as a consequence of such a theory
(Amelino-Camilia {\it et al.} 1998)\cite{am98}. A simpler formulation
for breaking LI by a small first order perturbation in the electromagnetic 
Lagrangian which leads to a renormalizable treatment has been given by
Coleman and Glashow (1999)\cite{co99}. Using this formalism, these authors 
have shown than only a very tiny amount of LI symmetry breaking is required 
to avoid the GZK effect by supressing photomeson interactions between
ultrahigh energy protons and the CBR. This LI breaking amounts to a 
difference of \cal{O}($10^{-23}$) between the maximum proton pion velocities. 
By comparison, Stecker and Glashow (2001) \cite{sg01} have placed an upper 
limit of \cal{O}($10^{-13}$) on the difference between the velocities of the 
electron and photon, ten orders of magnitude higher than required to eliminate
the GZK effect. 

\section{Is the GZK Effect All There Is?} 

There is a remaining ``dull'' possibility. Perhaps the GZK effect is  
consistent with the data and is all there is at ultrahigh energies.
The strongest case for trans-GZK physics comes from the AGASA results.
The AGASA group, which reported up to 17 events with energy greater than
or equal to $\sim$ 100 EeV (Sasaki, \etal
2001), \cite{sa01} has now lowered this number to 8 (see footnote 1). 
However, the HiRes Group have not confirmed the AGASA results, 
implying lower fluxes of cosmic rays above $\sim$ 100 EeV 
(T. Abu-Zayyad, \etal, in preparation; P. Sokolsky and E.C. Loh, 
private communication.) Even if the GZK effect is seen, top-down
scenarios predict the reemergence of a new component at even higher
energies (Aharonian, Bhattacharjee and Schramm 1992; Bhattacharjee and 
Sigl 2000). \cite{ah92} \cite{bh00}

The AGASA data indicate a significant deviation from pure GZK even if the 
source number is weighted like the local galaxy distribution (Blanton, 
\etal 2001) \cite{bl01} In addition to this discrepency, 
the fact that a flourescence detector, Fly's Eye, reported the 
highest energy event yet seen, {it viz.}, $E \simeq 300$ EeV,
makes the experimental situation interesting
enough to justify both more sensitive future detectors and the exploration
of new physics and astrophysics. 

\section{Signatures}

Future data which will be obtained with new detector arrays and satellites
(see section 9) will give us more clues relating to the origin of the
trans-GZK events by distinguishing between the various hypotheses which have
been proposed.

A zevatron origin (``bottom-up'' scenario) will produce air-showers 
primarily from primaries which are protons or heavier nuclei, with a much 
smaller number of neutrino-induced showers. The neutrinos will be 
secondaries from the photomeson interactions which produce the GZK effect 
(Stecker 1973; 1979; Engel, Seckel and Stanev 2001 and references therein).
\cite{st73} \cite{st79} \cite{en01} In addition, zevatron events may 
cluster near the direction of the sources.

A ``top-down'' (GUT) origin mechanism will not produce any heavier nuclei 
and will produce more ultrahigh energy neutrinos
than protons. This was suggested as a signature of
top-down models by Aharonian, Bhatacharjee and Schramm (1992). \cite{ah92} 
Thus, it will be important to look for the neutrino-induced air
showers which are expected to originate much more deeply in the atmosphere 
than proton-induced air showers and are therefore expected to be mostly
horizontal showers. Looking for these events can most easily be done with a
satellite array which scans the atmosphere from above (See Section 9.)

Top-down models also produce more photons than protons
However, the mean free path of these photons against
pair-production interactions with extragalactic low frequency radio photons
from radio galaxies is only a few Mpc at most (Protheroe and Biermann 1996).
\cite{pr96} The subsequent electromagnetic cascade and synchrotoron 
emission of the high energy electrons produced in the cascade dumps the 
energy of these 
particles into much lower energy photons (Wdowczyk, Tkaczyk and Wolfendale 
1972; Stecker 1973)\cite{wd72} \cite{st73}. However, the photon-proton ratio
is an effective tool for testing halo fragger models (See section 5.4.)

Another characteristic which can be used to distinguish between the 
bottom-up and top-down models
is that the latter will produce much harder spectra. If differential
cosmic ray spectra are parametrized to be of the form 
$F \propto E^{-\Gamma}$,
then for top-down models $\Gamma < 2$, whereas for bottom-up models
$\Gamma \ge 2$. Also, because of the hard source spectrum in the 
``top-down'' models, they should exhibit both a GZK suppression and a 
pileup just before the GZK energy.

If Lorentz invariance breaking is the explanation for the missing GZK 
effect, the actual absence of photomeson interactions should result the 
absence of a pileup effect as well.

\section{Present and Future Detectors}

\begin{figure}
\centerline{\psfig{figure=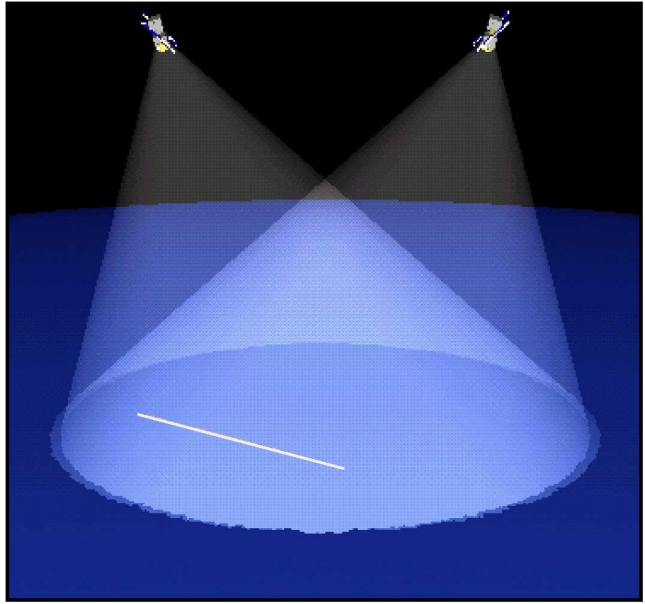,height=8cm}}

\caption{Two OWL satellites in low-Earth orbit observing the flourescent
track of a giant air shower. The shaded cones illustrate the field-of-view
for each satellite.}
\end{figure}

Of the ground-based ultra-high energy arrays, the AGASA array of particle
detectors in Japan is continuing to obtain data on ultrahigh energy cosmic 
ray-induced air showers. Its aperture is 200 km$^2$sr. 
The HiRes array is operating and will soon be publishing data. This array 
is an extension of the Fly's Eye which pioneered the technique of 
measuring the atmospheric fluorescence light in the near UV (300 - 400 nm 
range) that is
isotropically emitted by nitrogen molecules that are excited by the 
charged shower secondaries at the rate of $\sim$4 photons per meter per 
particle. Its estimated aperture is 1000 km$^2$sr at 100 EeV after inclusion
of a 10\% duty cycle (Sokolsky 1998) \cite{so98}. 

The southern hemisphere Auger array is expected to be on line in the near 
future. This will be a hybrid array which will consist of 1600 particle
detector elements similar to those at Havera Park and three or four 
flourescence detectors. Its expected aperture will be 7000 km$^2$sr for the
ground array above 10 EeV and $\sim$ 10\% of this number for the hybrid array.
The Telescope Array will will consist of eight separate flourescence detecting
telescope stations separated by 30 km. Its expected aperture will be 8000 
km$^2$sr with an assumed 10\% duty cycle. 

The next big step will be to orbit a system of space-based detectors which
will look down on the Earth's atmosphere to detect the trails of nitrogen
flourescence light made by giant extensive air showers.
The Orbiting Wide-angle Light collectors (OWL) mission is being proposed 
to study such showers from satellite-based platforms in low 
Earth orbit (600 - 1200 km). OWL would observe extended air showers 
from space via the air fluorescence technique, thus determining the 
composition, energy, and arrival angle 
distributions of the primary particles in order to deduce their origin.  
Operating from space with a wide field-of-view instrument
dramatically increases the observed target volume, and consequently the 
detected air shower event 
rate, in comparison to ground based experiments. The OWL baseline 
configuration will yield event rates that are more than two orders 
of magnitude larger than currently operating ground-based experiments.  
The estimated aperture for a two-satellite system is $2.5 \times 10^5$
km$^2$sr above a few tens of EeV after assuming a 10\% duty cycle.

Figure 9 illustrates two OWL satellites obtaining stereoscopic views of an 
air shower produced by an ultra-high energy cosmic ray. 
With an approximate 10\% duty factor, OWL will be capable of
making accurate measurements of giant air shower events with high statistics. 
It is expected to be able to detect more than 1000 showers per year with 
$E \ge $ 100 EeV (assuming an extrapolation of the
cosmic ray spectrum based upon the AGASA data). 

The European Space Agency is now studying the feasibility of placing such a 
light collecting detector on the International Space Station in order to 
develop the required technology to observe the flourescent trails of giant 
extensive air showers, to make such observations,
and to serve as a pathfinder mission for a later free flyer. This experiment
has been dubbed the Extreme Universe Space Observatory (EUSO) (see paper
of Livio Scarsi, these proceedings, for more details).
Owing to the orbit parameters and constraints of the International
Space Station, the effective aperture for EUSO will not be as large 
as that of a free flyer mission. 
A recent compendium of papers on observing giant air showers from space may be
found in Krizmanic, Ormes and Streitmatter (1998)\cite{kr98}.

\section{A Cloudy Present -- A Bright Future}

As of this writing, there is a disagreement in the trans-GZK event rate
between the AGASA and HiRes experimental groups. Thus, we are uncertain 
about the observational situation. The prospect of new physics and new
astrophysics at ultrahigh energies has produced a plethora of theoretical
ideas and papers. Indeed, if there are significant numbers of ultrahigh
energy events above 100 EeV, and especially above 300 EeV (which would rule
out the heavy nucleus scenario) many of the theoretical models presently 
proposed could be ruled out. This situation might then call for radically
new physics such as would involve violation of Lorentz invariance.

New and more powerful observational techniques are called for 
to obtain significantly large numbers of giant air shower events to analyse 
in order to accurately determine the flux and energy spectrum of trans-GZK 
cosmic rays. The Auger ground array is starting 
operation. New space experiments, EUSO and OWL, have been propsed. This author
hopes that they be built and flown and will provide the needed information.
Such experiments have the potential of breaking through to new insights about
the basic nature of the universe. 




\noindent Acknowledgements: I thank J. Krizmanic for his help on the 
future detectors section and for supplying Figs. 7 and 9 and R.
Streitmatter for reading over the manuscript.

{}


\begin{thebibliography}{} 


\bibitem[]{ab00} Abu-Zayyad, T. {\it et al.} 2000 \prl {\bf 84}, 4276.
\bibitem[]{afg02} Anchordoqui, L.A., Feng, J.L. and Goldberg, H. 2002,
{\it Phys. Lett. B} {\bf 535}, 302.
\bibitem[]{ah92} Aharonian, F., Bhattacharjee, P. and Schramm, D.N. 1992,
{\it Physical Review D} {\bf 46}, 4188.
\bibitem[]{ah02} Ahmad, Q.R., \etal 2002, \prl {\bf 89}, 011302.
\bibitem[]{ah02a} Ahn, E.-J., Cavaglia, M. and Olinto, A.V. 2002,
e-print hep-th/0201042.
\bibitem[]{al99} Alavi-Harati, A. {\it et al.} 1999 \prl {\bf 83}, 2128.
\bibitem[]{am98} Amilino-Camilia, G. {\it et al.} 1998 \nat {\bf 393}, 763.
\bibitem[]{an01} Anchordoqui, L.A., \etal 2001, {\it Physical Review D} {\bf
63}, 124009.
\bibitem[]{an02} Anchordoqui, L.A., \etal 2002, {\it Physical Review D} {\bf
65}, 124027.
\bibitem[]{ar99} Arkani-Hamed, N. Dimopoulos, S. and Dvali, G.R. 1999, {\it 
Physical Review D} {\bf 59}, 086004.
\bibitem[]{av00} Ave, M. {\it et al.} 2000, \prl {\bf 85}, 2244.
\bibitem[]{av02} Ave, M. {\it et al.} 2002, {\it Physical Review} {\bf 65},
063007.
\bibitem[]{} Barbot, C. \etal 2002, e-print hep-ph/0205230.
\bibitem[]{be88} Berezinsky, V.S. and Grigor'eva S.I. 1988, {\it Astronomy and 
Astrophysics} {\bf 199}, 1.
\bibitem[]{be97} Berezinsky, V., Kachelriess, M. and Vilenkin 1997, 
\prl {\bf 79}, 4302.
\bibitem[]{bh98} Bhattacharjee, P., Shafi, Q. and Stecker, F.W. 1998, \prl 
{\bf80}, 3698.
\bibitem[]{bh00} Bhattacharjee, P. and Sigl 2000 {\it Physics Reports}
{\bf327}, 109.
\bibitem{bi98} Biermann, P.L. 1998 in
{\it Workshop on Observing Giant Cosmic Ray Air Showers from Space}
ed. J.F. Krizmanic, J.F. Ormes  and R.E. Streitmatter, (New York: American 
Institute of Physics) p.22. 
\bibitem[]{bi87} Biermann, P.L. and Strittmatter, P.A. 1987, \apj {\bf 322}, 
643.
\bibitem[]{bi93} Bird, D.J. {\it et al.} 1993, \prl {\bf 71}, 3401.
\bibitem[]{bi94} Bird, D.J. {\it et al.} 1994, \apj {\bf 424}, 491.
\bibitem[]{bl01} Blanton, M., Blasi, P. and Olinto, A.V. 2001, {\it 
Astroparticle Physics} {\bf 15}, 275.
\bibitem[]{bl02} Blasi, P., Dick, R. and Kolb, E.W. 2002, {\it Astroparticle
Physics} {\bf 18}, 57.
\bibitem[]{bo99} Boldt, E. and Ghosh, P. 1999 \mnras {\bf 307}, 491.
\bibitem[]{bo00} Boldt, E. and Lowenstein, M. 2000, \mnras {\bf 316}, L29.
\bibitem[]{bo98} Bordes, J., {\it et al.} 1998, {\it Astroparticle Physics}
{\bf 8}, 135.
\bibitem[]{br73} Brown, R.W., {\it et al.} 1973, {\it Physical Review D} 
{\bf 8}, 3083.
\bibitem[]{bu98} Burdman, G., Halzen, F. and Ghandi, R. 1998 {\it Physics 
Letters B}
{\bf 417}, 107.
\bibitem[]{ch98} Cheung, D.J.H., Farrar, G.R. and Kolb, E.W. 1998, 
{\it Physical Review D} {\bf 57}, 4606.
\bibitem[]{ci99} Cillis, A.N. \etal 1992, {\it Physical Review D} {\bf 59}, 
113012.
\bibitem[]{cl00} Cline, D. and Stecker, F.W. 2000, {\it 1999 UCLA Workshop on 
Ultrahigh Energy Neutrino Astrophysics, science white paper} astro-ph/0003459.
\bibitem[]{co99} Coleman, S. and Glashow, S.L. 1999 {\it Physical Review D} 
{\bf 59} 116008.
\bibitem[]{da99} Dai, \etal 1999, \apj {\bf 511}, 739.
\bibitem[]{di82} Dimopoulos, S. Raby, S. and Wilczek, F. 1982, {\it Physics
Letters B} {\bf 112}, 133.
\bibitem[]{do88} Domokos, G. and Kovesi-Domokos, S. 1988 {\it Physical Review 
D} {\bf 38}, 2833.
\bibitem[]{do99} Domokos, G. and Kovesi-Domokos, S. 1999 \prl {\bf 82}, 1366.
\bibitem[]{do87} Domokos, G. and Nussinov, S. 1987 {\it Physics Letters B} 
{\bf 187}, 372.
\bibitem[]{dr94} Drury, L. 1994 {\it Contemporary Physics} {\bf35},232.
\bibitem[]{el95} Elbert, J.W. and Sommers, P. 1995, \apj {\bf 441}, 151.
\bibitem[]{en01} Engel, R., Seckel, D. and Stanev, T. 2001, {\it Physical
Review D} {\bf 64}, 093010.
\bibitem[]{er66} Erber, T. 1966, {\it Rev. Mod. Phys.} {\bf 38}, 626.
\bibitem[]{fa01} Falcke, H. 2001, {\it Reviews in Modern Astron.} {\bf 14}
15.
\bibitem[]{fa99} Fargion, D., Mele, B. and Salis, A. 1999 \apj {\bf 517}, 725.
\bibitem[]{fa96} Farrar, G.R. 1996 \prl {\bf 76}, 4111.
\bibitem[]{fa00} Farrar, G.R. and Piran, T. 2000  \prl {\bf 84}, 3527.
\bibitem[]{fe00} Fenimore, E.E. and Ramirez-Ruiz 2000, astro-ph/0004176.
\bibitem[]{ga00} Gaisser, T.K. 2000, in {\it Observing Ultrahigh Energy
Cosmic Rays from Space and Earth} CP566 (2000: American Inst. of Physics), 
pg. 566, e-print astro-ph/0011525.
\bibitem[]{ge74} Georgi, H. and Glashow, S.L. 1974, \prl {\bf 32} 438.
\bibitem[]{gl60} Glashow, S.L. 1960, {\it Nucl. Phys.} {\bf 22}, 579.
\bibitem[]{gi64} Ginzburg, V.L. and Ozernoi, L.M. 1964, {\it Zh. Eksp.
Teor. Fiz.} {\bf 47}, 1030.
\bibitem[]{go66} Gould, R,J. and Schreder, G.P. 1966, \prl {\bf 16}, 252.
\bibitem[]{gr66} Greisen, K. 1966 \prl {\bf 16}, 748.
\bibitem[]{ha02} Halzen, F. and Hooper, D. 2002, e-print astro-ph/0204527.
\bibitem[]{ha99} Hayashida, N., {\it et al.} 1999 {\it Astroparticle Physics}
{\bf 10}, 303.
\bibitem[]{hi85} Hill, C.T. and Schramm, D.N. 1985 {\it Physical Review D}
{\bf 31}, 564.
\bibitem[]{hi84} Hillas, A.M. 1984 {\it Annual Review of Astronomy and 
Astrophysics} {\bf 22}, 425.
\bibitem[]{ja00} Jain, P., {\it et al.} 2000 {\it Physics Letters B}
{\bf 484}, 267.
\bibitem[]{jo00a}Jones, F.C. 2000, paper presented at the {\it OWL Workshop, 
2000} (2000).
\bibitem[]{jo00} Jones, T.W. 2000, in {\it Proc. Seventh Taipei Workshop
on Astrophysics}, e-print astro-ph/0012483.
\bibitem[]{ka21} Kaluza, T. {\it Sitzungsberichten der Preussen Akademie
f\"{u}r Wissenschaften, Berlin (Math. Phys.)} {\bf K1}, 966.
\bibitem[]{ki76} Kibble, T.W.B. 1976 {\it Journal of Physics A} {\bf9}, 1387.
\bibitem[]{kl26} Klein, O. 1926, {\it Zeitschrift f\"{u}r Physik} {\bf 37}, 
895.
\bibitem[]{kr98} Krizmanic, J.F. Ormes, J.F. and Streitmatter, R.E. 1998  
{\it Workshop on  Observing Giant Cosmic Ray Air Showers from Space}
(New York: American Institute of Physics).
\bibitem[]{kr99} Krolik, J. 1999, \apj {\bf 515}, L73.
\bibitem[]{ku98} Kuz'min, V.A., and Rubakov, V.A. 1998, {\it Phys. Atom Nucl.}
{\bf 61}, 1028.
\bibitem[]{la53} Landau, L.D. and Pomeranchuk, I.J. 1953, {\it Doklady
Acad. Nauk SSSR} {\bf 92}, 535.
\bibitem[]{li63} Linsley, J. 1963, {\prl},{\bf 10}, 146.
\bibitem[]{ma94} Ma, C.-P. and Bertschinger, E. 1994  \apj {\bf 434}, L5.
\bibitem[]{ma98} Mao, S. and Mo, H.J. 1998, {\it Astronomy and Astrophysics} {\bf 339}, L1.
\bibitem[]{mat94} Mather, J. {\it et al.} 1994, \apj {\bf 420}, 439.
\bibitem[]{mi56} Migdal, A.B. 1956, {\it Physical Review} {\bf 103}, 1811.
\bibitem[]{na00} Nagano, M. and Watson, A.A. 2000 {\it Reviews of Modern 
Physics} {\bf 72}, 689.
\bibitem[]{no02} Norris, J. 2002, \apj , in press, e-print astro-ph/0201503.
\bibitem[]{nu99} Nussinov, S. and Schrock, R. 1999, {\it Physical Review D}
{\bf 59}, 105002. 
\bibitem[]{ol00} Olinto, A 2000, in {\it Observing Ultrahigh Energy
Cosmic Rays from Space and Earth} CP566 (2000: American Inst. of Physics), 
pg. 99, e-print astro-ph/0011106.
\bibitem[]{pr96} Protheroe, R.J. and Biermann, P.L. 1996, {\it Astroparticle 
Physics} {\bf 6}, 45.
\bibitem[]{pu76} Puget, J.L., Stecker, F.W. and Bredekamp, J. 1976, \apj 
{\bf 205}, 638.
\bibitem[]{pe65} Penzias, A.A. and Wilson, R.W. 1965 \apj {\bf 142}, 419.
\bibitem[]{pr96} Protheroe, R.J. and Biermann, P.L. 1996, {\it Astroparticle
Physics} {\bf 6}, 45. 
\bibitem[]{ra99} Randall, L. and Sundrum, R., \prl {\bf 83}, 3370.
\bibitem[]{ro88} Robinett, R.W. 1988, {\it Physical Review D} {\bf 37}, 84.
\bibitem[]{sa68} Salam, A. 1968, in {\it Elementary Particle Theory (Proc. 
8th Nobel Symp.)} ed. N. Svartholm (Stockholm: Almqvist and Wiksells Vorlag), 
pg. 367. 
\bibitem[]{} Sarkar, S. and Toldr\`{a}, R. 2002, {\it Nucl. Phys. B}
{\bf 621}, 2002.
\bibitem[]{sa01} Sasaki, N. \etal 2001, {\it Proc. 27th Intl. Cosmic Ray
Conf., Hamburg} {\bf 1}, 333.
\bibitem[]{sa72} Sato, H. and Tati, T. 1972, {\it Progress in Theoretical 
Physics} {\bf 47}, 1788.
\bibitem[]{sc99} Schmidt, M. 1999, \apj {\bf 523}, L117.
\bibitem[]{sc89} Schaefer, R.K., Shafi, Q. and Stecker, F.W. 1989, \apj 
{\bf 347}, 575.
\bibitem[]{sh84} Shafi, Q. and Stecker, F.W. 1984, \prl {\bf 53}, 1292.
\bibitem[]{sh02} Shinozaki, {\it et al.} 2002, \apj {\bf 571}, L117.
\bibitem[]{sc02} Scully, S.T. and Stecker, F.W. 2002 {\it Astroparticle 
Physics} {\bf 16}, 271.
\bibitem[]{sm92} Smoot, G. {\it et al.} 1992, \apj {\bf 396}, L1.
\bibitem[]{so98} Sokolsky, P. 1998, in {\it Workshop on Observing Giant 
Cosmic Ray Air Showers from Space} ed. J.F. Krizmanic, J.F. Ormes and R.E. 
Streitmatter, (New York: American Institute of Physics) p. 65. 
\bibitem[]{st95} Stanev, T. {\it et al.} 1995 \prl {\bf 75}, 3056.
\bibitem[]{st68} Stecker, F.W. 1968, \prl {\bf 21}, 1016. 
\bibitem[]{st73} Stecker, F.W. 1973, {\it Astrophysics and Space Science}
{\bf 20}, 47.
\bibitem[]{st79} Stecker, F.W. 1979, \apj {\bf 228}, 919.  
\bibitem[]{st89} Stecker, F.W. 1989, {\it Nature} {\bf 342}, 401.
\bibitem[]{st00} Stecker, F.W. 2000, {\it Astroparticle Physics} {\bf 14}, 207.
\bibitem[]{sg01} Stecker, F.W. and Glashow, S.L. 2001, {\it Astroparticle 
Physics} {\bf 16}, 97.
\bibitem[]{st99} Stecker, F.W. and Salamon, M.H. 1999, \apj {\bf 512}, 521.
\bibitem[]{st91} Stecker, F.W. {\it et al.} 1991, \prl {\bf 66}, 2697.  
\bibitem[]{ta98} Takeda, M. {\it et al.} 1998, \prl {\bf 81}, 1163.
\bibitem[]{ta99} Takeda, M. {\it et al.} 1999 \apj {\bf 522}, 225.
\bibitem[]{tr79} Tremaine, S. and Gunn, J.E. 1979, \prl {\bf 42}, 407.
\bibitem[]{ty01} Tyler, C., Olinto, A.V. and Sigl, G. 2001, {\it Physical 
Review D} {\bf 63}, 055001.
\bibitem[]{vi95} Vietri, M. 1995 \apj {\bf 453}, 883.
\bibitem[]{wa95} Waxman, E. 1995, \prl {\bf 75}, 386.
\bibitem[]{wd72} Wdowczyk, J., Tkaczyk, T. and Wolfendale, A.W. 1972 
{\it Journal of Physics A} {\bf 5}, 1419.
\bibitem[]{we82} Weiler, T.J. 1982, \prl {\bf 49}, 234.
\bibitem[]{we99a} Weiler, T.J. 1999 {\it Astroparticle Phys.} {\bf 11}, 303.
\bibitem[]{we67} Weinberg, S. 1967, \prl {\bf 19}, 1264.
\bibitem[]{we76} Weinberg, S. 1979, \prl {\bf 42}, 850. 
\bibitem[]{we99} Weinheimer, C., {\it et al.} 1999, {\it Phys. Lett. B}
{\bf 460}, 219.
\bibitem[]{wr92} Wright, E.L. 1992 \apj {\bf 396}, L13.
\bibitem[]{za66} Zatsepin, G.T. and Kuz'min, V.A. 1966, {\it Zh. Esks. Teor. 
Fiz., Pis'ma Red.} {\bf 4}, 144.
\bibitem[]{}


\end{thebibliography}
\end{document}